\documentstyle[aps,multicol,prb,epsf]{revtex}
\begin{document}
\title{Magnetothermopower and magnon-assisted transport in ferromagnetic tunnel
junctions}
\author{Edward McCann and Vladimir I. Fal'ko}
\address{Department of Physics, Lancaster University,\\
Lancaster, LA1 4YB, United Kingdom}
\date{\today}
\maketitle

\begin{abstract}
{We present a model of the thermopower in a mesoscopic tunnel junction
between two ferromagnetic metals based upon magnon-assisted tunneling processes.
In our model, the thermopower is generated in the course of thermal
equilibration between two baths of magnons, mediated by electrons.
We predict a particularly large thermopower
effect in the case of a junction between two half-metallic ferromagnets
with antiparallel polarizations, $S_{AP} \sim - (k_B/e)$, in contrast to
$S_{P} \approx 0$ for a parallel configuration. }
\end{abstract}


\begin{multicols}{2}

Spin valve systems and magnetic multilayers displaying giant magnetoresistance
effects also exhibit substantial magnetothermopower
\cite{sak91,con91,pir92,avdi93,shi93,sat98} with a strong temperature dependence.
In metals, the thermopower $S$ is related to the conductivity of electrons
taken at a certain energy, $\sigma (\epsilon )$, by the Mott formula,\cite{zim64}
$S=-(\pi ^{2}k_{B}^{2}T/3e)\left( \partial \ln \sigma (\epsilon
)/\partial \epsilon \right) _{\epsilon _{F}}$, so that it typically
contains a small parameter such as $k_{B}T/\epsilon_{F}$.
Theories of transport in magnetic multilayers with highly transparent
interfaces based upon the use of the Mott formula have explained the difference
between thermopower in the parallel (P) and anti-parallel (AP)\ configuration
of ferromagnetic layers as due to either the
difference in the energy dependence of the density of states for
majority and minority spin bands in ferromagnetic layers,\cite{shi96,tsy99}
or a different efficiency of electron-magnon scattering for carriers in
opposite spin states.\cite{pir92}
In particular, the electron-magnon interaction in a ferromagnetic layer
was incorporated to explain
the observation \cite{pir92} of a
strong temperature dependence of $S(T)$ and gave, theoretically, a
much larger thermopower in the parallel configuration of multilayers with
highly transparent interfaces than in the antiparallel one, $S_{P}\gg S_{AP}$.

In this paper we investigate a model of the electron-magnon interaction
assisted thermopower in a mesoscopic size ferromagnet/insulator/ferromagnet
tunnel junction, which yields a different prediction.
In the model we study below,
the bottle-neck of both charge and heat transport lies in a
small-area tunnel contact between ferromagnetic metals held at different
temperatures, $T\pm \Delta T/2$.
The thermopower is generated in the course of thermal
equilibration between two baths of magnons, mediated by electrons.
We find that the magnetothermopower effect is most pronounced in the case
of half-metallic ferromagnets, where the exchange spin splitting $\Delta$
between the majority and minority conduction bands is greater
than the Fermi energy $\epsilon _{F}$ measured from the bottom of the
majority band, and the Fermi density of states in the minority band is zero. In
a highly resistive antiparallel configuration of such a junction, where the
emission/absorption of a magnon would lift the spin-blockade of electronic
transfer between ferromagnetic metals, we predict a large thermopower
effect, whereas in the lower-resistance parallel configuration thermopower
appears to be relatively weak:
\begin{equation}
S_{AP}\approx -0.64\frac{k_{B}}{e};\;\;
\frac{S_{P}}{S_{AP}}\sim \frac{k_{B}T}{\epsilon_{F}}.
\label{MainResHM}
\end{equation}
We also found that for a junction between two conventional ferromagnetic
metals, the ability of electronic transfer assisted by magnon
emission/absorption to create thermopower depends on the difference between
the size of majority/minority band Fermi surfaces and it is reduced by a temperature
dependent factor $g(T) \sim (k_{B}T/\omega _{D})^{3/2}$.
The latter reflects the fractional change in the net
magnetization of the reservoirs due to thermal magnons (Bloch's $T^{3/2}$ law).

Below, we describe the calculation of the thermopower for the case
of a tunnel contact between two half-metallic ferromagnets
and, then, we present its generalization to conventional ferromagnetic metals.
We obtain an expression for the current $I(V,\Delta T)$ between bulk
ferromagnetic reservoirs, as a function of bias voltage, $V$, and of the
temperature drop, $\Delta T$, and, then, determine the thermopower
coefficient $S=-V/\Delta T$ by satisfying the relation $I(V,\Delta T)=0$.
The expression for the current was derived using the balance equation,
which takes into account competing elastic and inelastic electron transfer
processes across the tunnel junction.

Let us consider, first, the AP configuration of ferromagnetic electrodes,
with spin-$\uparrow$ majority electrons on the left hand side of the
junction and spin-$\downarrow$ on the right.
For such an alignment, elastic tunneling of carriers between electrodes
is blocked by the absence of available states for a spin-polarized electron
on the other side of an insulating barrier,
whereas electron transfer may happen via tunneling processes
assisted by a simultaneous emission/absorption of a magnon.\cite{Edwards}
Since tails of wavefunctions of majority-spin ($\uparrow $) electrons
close to the Fermi level on the left hand side penetrate into the forbidden
region on the right, an electron on one side of the junction acquires a weak
coupling with core magnetic moments (and, therefore, magnons) on the other
side. A characteristic event can be viewed as a two-step quantum process.
First, an electron tunnels into a virtual intermediate high-energy state in the
minority band. Then, it incorporates itself into the majority band by
flipping spin in a magnon-emission process.
Following the tunneling Hamiltonian approach,\cite{mah90} the amplitude for a
spin-$\uparrow $ electron with wave number ${\bf k}$ on the left to finish in a
state $(\downarrow , {\bf k^{\prime}})$ on the right
after emitting a spin-wave with wavenumber ${\bf q}$ can be estimated using
second order perturbation theory with
respect to the electron - magnon interaction and the tunneling matrix element
$t_{{\bf k}, {\bf k^{\prime} + q}}$:
\begin{equation}
A_{{\bf k},{\bf k^{\prime} + q}} =
\frac{t_{{\bf k}, {\bf k^{\prime} + q}} \, \Delta}
{\sqrt{2\xi {\cal N}}
\left( \Delta +\epsilon _{{\bf k^{\prime}+q}}-\epsilon_{\bf k} \right)}
\approx \frac{t_{{\bf k},{\bf k^{\prime} + q}}}{\sqrt{2\xi {\cal N}}},
\label{probability}
\end{equation}
For $k_{B}T,eV \ll \Delta$, when both initial and final electron states should
be taken close to the Fermi level, only long wavelength magnons can be emitted,
so that the energy deficit in the virtual states can be approximated as
$\Delta +\epsilon _{{\bf k^{\prime }+q}}-\epsilon_{\bf k} \approx \Delta$.
As noticed in Refs.~\onlinecite{Edwards,m+f01}, this cancels out the large
exchange parameter since the same electron-core spin exchange constant appears both in the
splitting between minority and majority bands and in the electron-magnon
coupling. [The number of localized moments in a ferromagnet ${\cal N}$
appears in Eq. (\ref{probability}) as we normalize both electron and magnon
plane waves to the system volume, and $\xi $ is spin per unit magnetic cell.]

\begin{figure}
\epsfxsize=\hsize
\epsffile{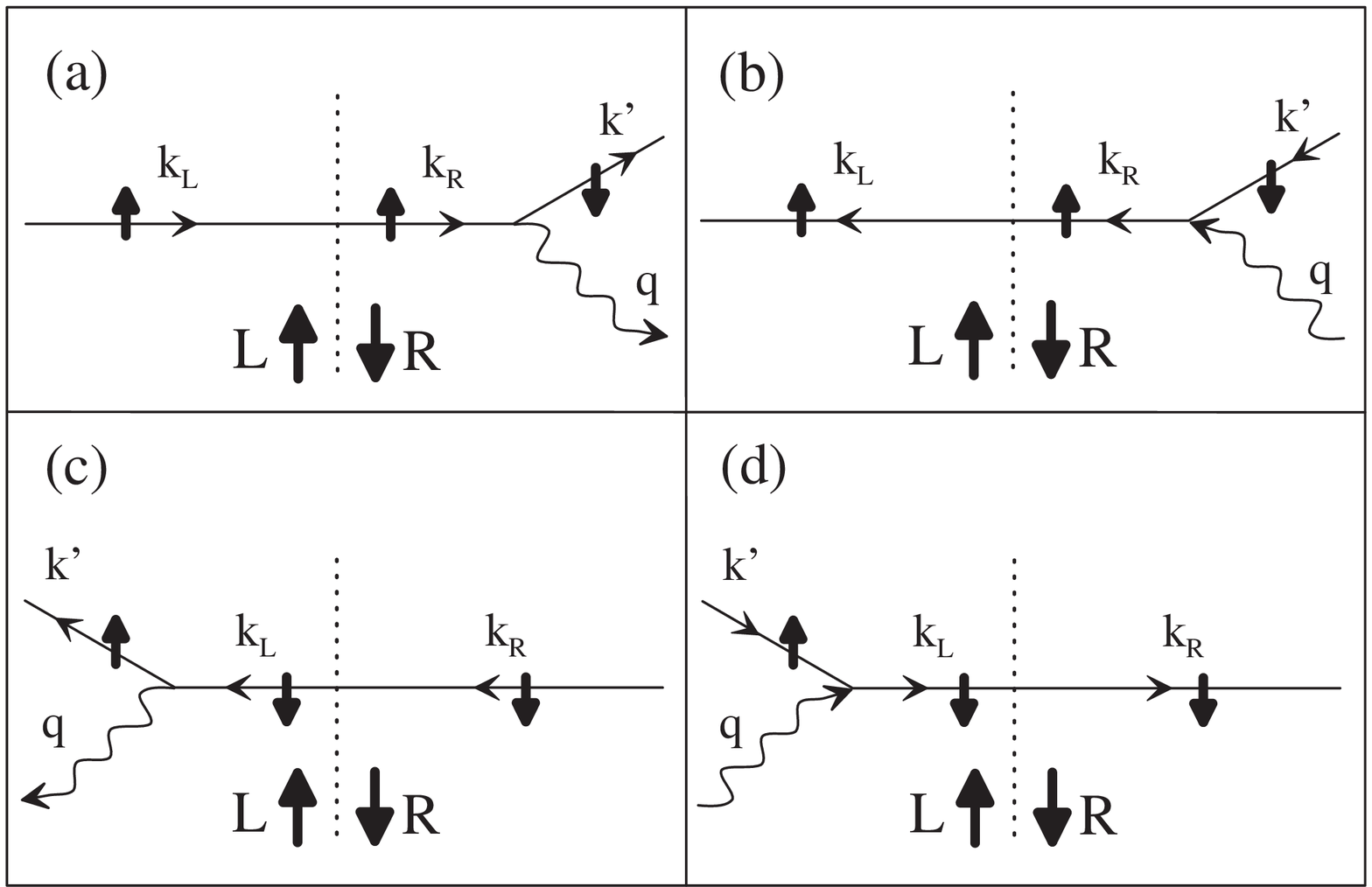}
{\setlength{\baselineskip}{10pt} FIG.\ 1.
Schematic of magnon-assisted tunneling across a junction with half-metallic
electrodes in the anti-parallel configuration.
Four processes which, to lowest order in the electron-magnon interaction,
contribute to magnon-assisted tunneling.
(a) and (c) involve magnon emission on the right and left hand sides,
respectively, whereas (b) and (d) involve magnon absorption
on the right and left.}
\end{figure}

To complete the balance equation describing electron transfer between
half-metallic electrodes, one has to take into account four magnon-assisted
tunneling processes depicted in Figure~1. Below, we describe them in detail
assuming that the tunnel barrier is flat, so that the parallel component of
the electron momentum conserves upon tunneling.
Two of these processes, (a) and (b), involve the interaction of electrons
with a thermal bath of magnons on the right hand side of the junction and
are responsible for transferring electrons in opposite directions.
The process (a) begins with a $\uparrow$ electron on the left with wavevector
${\bf k}_{{\rm L}}{\bf =(k}_{{\rm L}}^{\Vert },k_{{\rm L}}^{z})$
[with occupation number $n_{{\rm L}}({\bf k}_{{\rm L}})$],
which tunnels through the barrier into an intermediate virtual $\uparrow$ state
on the right
(${\bf k}_{{\rm R}}{\bf =(k}_{{\rm L}}^{\Vert },k_{{\rm R}}^{z})$).
Then, this electron flips spin by emitting a magnon with
wavevector ${\bf q}$
[this process is stimulated by the occupancy factor of
thermal magnon excitations $1+N_{{\rm R}}({\bf q})$],
and, thus, incorporates itself into the majority spin band on
the right, provided the final $\downarrow $ state
(${\bf k^{\prime }}={\bf k}_{{\rm R}}-{\bf q}$) is not occupied [which has
probability $1-n_{{\rm R}}({\bf k}_{{\rm R}}-{\bf q})$].
The process (b) is the reverse to the process (a).
It begins with a $\downarrow$ electron on the
right with wavevector ${\bf k^{\prime }}={\bf k}_{{\rm R}}-{\bf q}$, that absorbs
a magnon, flips its spin and moves into a virtual minority-spin state on the right.
Then, it tunnels into an empty final state in the majority spin band in the
left reservoir.
The balance between these two processes contributes to the total current as
\begin{eqnarray}
I_{{\rm ab}} &=&-4\pi ^{2}\frac{e}{h}\int_{-\infty }^{+\infty }\!\!d\epsilon
\sum_{{\bf k}_{{\rm L}}{\bf k}_{{\rm R}}{\bf q}}
\left| A_{ {\bf k}_{{\rm L}}, {\bf k}_{{\rm R}} }\right| ^{2}
\nonumber \\
&&\times \,\delta (\epsilon -\epsilon _{{\bf k}_{{\rm L}}})\,\delta
(\epsilon -eV-\epsilon _{{\bf k}_{{\rm R}}-{\bf q}}-\omega _{{\bf q}})
\nonumber \\
&&\times \left\{ n_{{\rm L}}({\bf k}_{{\rm L}})\left[ 1-n_{{\rm R}}
({\bf k}_{{\rm R}}-{\bf q})\right] \left[ 1+N_{{\rm R}}({\bf q})\right] \right. - 
\nonumber \\
&&\;\;\;\;-\left. \left[ 1-n_{{\rm L}}({\bf k}_{{\rm L}}{\bf )}\right]
n_{{\rm R}}({\bf k}_{{\rm R}}-{\bf q})N_{{\rm R}}({\bf q})\right\} ,
\label{iab}
\end{eqnarray}
where
$n_{{\rm L/R}}({\bf k})=[\exp \{(\epsilon _{{\bf k}}-\epsilon _{{\rm F}
}^{{\rm L/R}}\}/k_{B}T_{{\rm L/R}})+1]^{-1}$,
$\epsilon _{{\rm F}}^{{\rm L}}-\epsilon _{{\rm F}}^{{\rm R}}=-eV$,
$N_{{\rm L/R}}({\bf q})=[\exp (\omega _{{\bf q}}/k_{B}T_{{\rm L/R}})-1]^{-1}$,
and
$T_{{\rm L/R}}=T\pm \Delta T/2$.

Two other processes shown in Figure~1(c) and (d) involve emission/absorption of
magnons on the left hand side of the junction.
Their contribution to the total current is 
\begin{eqnarray}
I_{{\rm cd}} &=&-4\pi ^{2}\frac{e}{h}\int_{-\infty }^{+\infty }\!\!d\epsilon
\sum_{{\bf k}_{{\rm L}}{\bf k}_{{\rm R}}{\bf q}}
\left| A_{ {\bf k}_{{\rm L}}, {\bf k}_{{\rm R}} }\right| ^{2}
\nonumber \\
&&\times \,\delta (\epsilon -eV-\epsilon _{{\bf k}_{{\rm R}}})\,\delta
(\epsilon -\epsilon _{{\bf k}_{{\rm L}}-{\bf q}}-\omega _{{\bf q}})
\nonumber \\
&&\times \left\{ -n_{{\rm R}}({\bf k}_{{\rm R}})\left[ 1-n_{{\rm L}}
({\bf k}_{{\rm L}}-{\bf q})\right] \left[ 1+N_{{\rm L}}({\bf q})\right] +\right.
\nonumber \\
&&\left. \;\;\;\;+\left[ 1-n_{{\rm R}}({\bf k}_{{\rm R}})\right]
n_{{\rm L}}({\bf k}_{{\rm L}}-{\bf q})N_{{\rm L}}({\bf q})\right\} .
\label{icd}
\end{eqnarray}
After combining them together into an expression for the total current
$I=I_{ab}+I_{cd}$, and, then, performing summation over wave numbers and
integration over initial electron energies, we arrived at the following
expression
\begin{eqnarray}
I &=& + \frac{3}{4}
\left( \frac{G_P}{\xi e} \right)
\left( \frac{k_BT}{\omega_D} \right)^{3/2}
\left[ a \, eV - b \, k_B \Delta T \right]  ,  \label{itot}
\end{eqnarray}
where
$a = 3\Gamma (3/2) \zeta (3/2)$,
$b = (5/2)\Gamma (5/2) \zeta (5/2)$,
$\Gamma (x)$ is the gamma function, and $\zeta (x)$ is Riemann's zeta
function.
Here, all properties of the interface are incorporated into a single parameter
$G_P$ which coincides with the linear conductance of the same mesoscopic junction
in the P configuration.
For a flat, clean barrier of area $A$,
where the parallel component of momentum is conserved upon tunneling,
we consider the tunneling matrix element to have the form
$t_{{\bf k}, {\bf k^{\prime}}} =
\delta_{{\bf k_{\Vert }}, {\bf k_{\Vert }^{\prime}}} \, t \,
| h^2 v_L^z v_R^z/L^2|^{1/2}$,
which gives
$G_P \approx 4 \pi^{2} (e^2/h) \left| t \right|^{2} (A\Pi_{+}/h^{2})$
where $v_{L(R)}^z$ is the perpendicular component of velocity on the left (right)
side, $L$ is the length of an electrode, $t$ is the barrier transparency,
and $\Pi_{+}$ is the area of the maximal cross-section of the Fermi surface
of majority electrons in the plane parallel to the interface.
When deriving Eq.~(\ref{itot}), we also assumed a quadratic magnon
dispersion, $\omega _{q}=Dq^{2}$, and $k_B T \ll \omega_D$,
where $\omega _{D}=D(6\pi^{2}/v)^{2/3}$ is the Debye magnon energy,
and $v$ is the volume of a unit cell.

The thermopower coefficient $S = - V/\Delta T$ can be found by setting the
total current in Eq. (\ref{itot}) to zero and determining the voltage
created by the temperature difference.
As a result, the tunneling conductance $G_P$ cancels from the final answer,
and, in the antiparallel configuration,
$S_{AP}\approx -0.64k_{B}/e$.\cite{noteAF}
In contrast to the AP configuration,
magnon-assisted tunneling cannot contribute to the electron transfer between
two electrodes in the P configuration, since both initial and final electron states
should have the same spin polarization in order to belong to the majority bands
in both of the reservoirs.
As a result, the linear conductance of such a junction is
formed without the involvement of magnon-assisted processes, and the thermopower
may only appear due to the energy-dependent electron tunneling
density of states, having the order of magnitude of $S_{P}\sim
(k_{B}/e)(k_{B}T/\epsilon _{F})$.

A generalization to conventional ferromagnetic metals of the proposed theory of
the magnon-assisted (ma) tunneling contribution to the thermopower yields
\begin{equation}
S_{AP}^{{\rm ma}} = - ( k_{B}/e) \, g \,\theta ;
\quad S_{P}^{{\rm ma}}=0,
\label{patialSP}
\end{equation}
\[
g \approx \frac{1.7}{\xi }\left( \frac{k_{B}T}{\omega _{D}}\right)
^{3/2};\;\theta =\left\{
\begin{array}{l@{\quad \quad\!\!\!\!\!}l}
(\Pi_{+}-\Pi_{-})/\Pi_{-}\,, & {\rm flat} \\ 
(\Pi_{+}^{2}-\Pi_{-}^{2})/(\Pi_{+}\Pi_{-}), & {\rm diff}
\end{array}
\right.
\]
where $\Pi _{\pm }$ is the area of the maximal cross-section of the Fermi
surface of majority/minority electrons in the plane parallel to the
interface ($\Pi _{+}>\Pi _{-}$), $\xi $ is the spin of localized
moments, and $\omega _{D}$ is the magnon bandwidth.
The function $g(T)$ is proportional to the fractional change in the net
magnetization due to thermal magnons (Bloch's $T^{3/2}$ law) and the
function $\theta$ is written for both a flat, clean interface (`flat')
and a diffusive tunnel barrier (`diff').
This result was obtained after some amendments to the above analysis were made.
First, the linear conductance in the AP configuration is not suppressed
because an elastic tunneling channel is opened between the majority band on one side
and the minority band on the other, which reduces the thermopower.
Secondly, for the AP configuration, in addition to the magnon-assisted tunneling processes
that enable transitions from majority initial to majority final states
via an intermediate minority state
(as described already for the half-metallic case and shown in Fig.~1),
one should take into account the possibility of magnon-assisted tunneling processes
that enable transitions from minority initial to minority final states
via an intermediate majority state.
A transition via a majority (minority) intermediate state results in the transfer
of electrons in the same (opposite) direction as the net polarization transfer
between two baths of magnons so that
the additional processes partially compensate the thermally excited currents.

The authors thank G.~Tkachov and A.~Geim for discussions. This work was
supported by EPSRC.

\end{multicols}


\begin{references}
\bibitem{noteAF}
The sign of the thermopower coefficient is
specified for the electron\ (charge $-e$) transfer process and under
the assumption that the exchange between conduction band and core
electrons has a ferromagnetic sign.
For antiferromagnetic exchange, the sign of the thermopower would be opposite.
One can verify this statement by taking into account that processes,
described for example in Fig.~1(a) and (b), would
involve magnons on the opposite side of the junction,
hence the current $I_{{\rm ab}}$ would be determined by magnon occupation
numbers $N_{{\rm L}}({\bf q})$.
\end{references}
\end{document}